# Performance Modeling and evaluation of Traffic management for Mobile Networks by SINR Prediction

K.K.Guatam
Department of Computer Science & Engineering
Roorkee Engineering & Management Technology Institute
Shamli(247774) India
-

Anurag Rai
Department of Information Technology
College of Engineering Roorkee
Roorkee (247667) India
-

*Abstract*— **Over the recent years a considerable amount of effort has been devoted towards the performance evaluation and prediction of Mobile Networks. Performance modeling and evaluation of mobile networks are very important in view of their ever expending usage and the multiplicity of their component parts together with the complexity of their functioning. The present paper addresses current issues in traffic management and congestion control by (singal-to-interference-plus-noise ratio) SINR prediction congestion control, routing and optimization of cellular mobile networks.**

## I. INTRODUCTION

Over the recent years a considerable amount of effort has been devoted towards the performance, evaluation of wireless mobile networks (WMN). A considerable amount of research has been used to characterize user and calling behavior and their performance impact on wireless mobile networks. At present the mobility in most mobile networks is confined to the end users only.

With the development of mobile compression, the CAC schemes are generally adopted by setting thresholds for hand off calls and new call differently given the traffic condition and it is the maximum number of users that can be supported.[1,2] In realistic systems, information about the traffic management quality cannot; be instantaneous but is outdated to some degree. Firstly, the SINR estimation in the receiver takes some time and secondly, the user has to wait for the next channel allocation to report his SINR to the base station. The system may need to block incoming users if all of the entire band width has been used up to provide the highest QoS to existing users. However, if these existing users can be degraded to a lower but acceptable QoS level, it is possible to reduce the blocking probability without degrading the QoS of existing users. A graceful degradation mechanism is proposed in (1). Thus a system could free some bandwidth allocation for new users. In this paper, we address current issues in traffic management of cellular mobile networks by the use of SINR prediction, the SINR is calculated for mobile user equipments in every transmission time interval for the traffic management of mobile networks. In traffic management and congestion control, courcoubetis and series device new procedures and tools for the analysis of network traffic measurement.

## II. MODEL DESCRIPTION

We consider uplink communication in a wireless mobile network. As an accepted call does not always send data frames. then for best traffic, we consider the activity factor $\ell$ as the probability that a call is active. We represent QoS requirement of traffic by required transmission rate. The required transmission rate can be obtained by setting the target level.

Often these intra–and inter–traffic interferences of calls can be large so that the target bit error rate of traffic interferences(BERIT) cannot be achieved temporarily, which is called outage.[3] The outage probability needs to approach zero (as close as possible) and can be different for each class. Here we assume for traffic management the allowed outage probability is the same.

## III. OUTAGE PROBABILITY FOR TRAFFIC

In a mobile network, a traffic management that supports a single class of calls, the outage probability is given by [2,4].
$$Pout = Pr \{N^a + M^a > (3/2) G(x^{-1} - (Yb/N0^{j1})^{+1}\}\ldots\ldots(1)$$
When $N^a$, $M^a$, G, X, $Y^b$, and N0 represent the number of active calls in the current call. Similarly, in a network that support L-Class of calls, we obtain

$$P^j out = Pr \{ \sum_{i=1}^{L} (Ybi / Ybj) Ci (Ni^a + Mi^a) \geq Aj\} = Pr$$

$$\{ \sum_{i=1}^{L} \theta I (Ni^a + Mi^a) \geq \eta j\}\ldots\ldots\ldots (2)$$

When i, j represent traffic call classes (TCC), Ci is the number of orthogonal codes needs for a TCC, 'i'. By the Gaussian random variable from the limit theorem and we can write control the outage probability of a TCC' 'j', as

$P^j out$ = Q ( $\eta$ j - $\lambda / \partial \lambda$ )……..(3)

Where Q ( $\xi$ ) = 1/ $\sqrt{}$ 2 $\pi$ $\int_{\xi}^{\infty}$ e$^{-x2/2}$ dx

And represent the total traffic receive single power (TRSP) i.e.

$\sum_{i=1}^{L} \theta$ I (Ni$^a$+Mi$^a$)

Therefore $\overline{\lambda}$ = (1+f1) $\sum_{i=1}^{L} \theta$ I $\overline{N}$ $^a$i,…… (4)

And $\partial^2 \lambda$ = $\sum_{i=1}^{L} \theta^2$ I ( $\partial$ i$^2$+f2 $\overline{N}$ $^a$i)

Where $\overline{N}$ $^a$i and $\partial^2$I indicate the mean and variance of N$^a$ I.
According to the assumption of TCC equal outage probability for each class, $\eta$ I = $\eta$ j for all I and j, therefore TCC received single power meets the following relation.

$\theta$ I / $\theta$ j = CiXi (3G - 2CjXj) $\diagup$ CjXj(3G+ 2CiXj)…....(6)

This indicates that the power allocation refers to the target of TCC outage probability Call Admission Control (CAC):

## IV. SYSTEM MODEL

The Communication system under consideration can be defined as r[k] = $\sum_{i=0}^{L}$ h [l]& [k-l] + z [k] ……….(7)

Where r [k] received call sequence h [l] unknown channel for traffic with memory L', z[k] is an independent and identically distributed Gaussian notice sequence. [5,6]

Then traffic management symbol sequence s [k] is drawn from M-ary alphabet, A with equal probability, the vector version of (1) can be written as

$$\begin{bmatrix} r[k] \\ : \\ : \\ r[1] \\ r[0] \end{bmatrix} = \begin{bmatrix} S[K-L] & ... & S[K] \\ : & ... & : \\ : & ... & : .. \\ S[1-L] & ... & S[1] \\ S[-L] & ... & S[0] \end{bmatrix} \begin{bmatrix} h[l] \\ : \\ : \\ h[1] \\ h[0] \end{bmatrix} + \begin{bmatrix} z[k] \\ : \\ : \\ z[1] \\ z[0] \end{bmatrix}$$

Where Sk is toeplitz data matrix.

## V. CALL ADMISSION CONTROL FOR TRAFFIC MANAGEMENT

I call Admission Control for traffic Management, (CACTM) the outage Probability is very small, defined as $\frac{\partial Q(\eta)}{\partial \eta} < 0$ we can show that $\frac{\partial P_{out}}{\partial N_i} = \alpha_i \frac{\partial P_{out}}{\partial N_i} > 0$ where $\alpha_i$ is the active factor for (CACTM) a class I call. It is clear that the average rate for mobile network (ARRMN) and outage probability increase with the number of users.

## VI. NUMERICAL RESULT

We now compare the performance of the two CACs through numerical analysis. [7] The system bandwidth is 2.50MHz and each code can carry information bits at the rate of 19.2(kbps) so that the processing gain is 256. Two types of calls are considered to manifest the effect of traffic parameters on performance. Class 1 and 2 calls are voice traffic and we set their transmission rates after channel coding at 19.29(kbps). They have different Mobile Network Average Revenue Rate (MNARR) for the traffic management requirement of less than 10$^{-4}$ and 10$^{-6}$, respectively, and their activity factors are set at 1.0. The coefficient for intercall interference modeling are chosen as f1 = 0.114 and f2 = 0.44(12).

## VII. FRAME WORK

*Angle – SINR Table:-*

In order to make the directional routing effective for call admission control system, a node should know how to set its transmission direction effectively to transmit a packet to its neighbors. So each node periodically collects its neighborhood information and forms an Angle- SINR Table (AST). Sinu$^s$ m(t) (Signal – to – Interference and Noise Ratio) is a number associated with each link 1$^u$ n, m, and is a measurable indicator of the strength of radio connection from node n to node m at an angle u with respect to n and as perceived by m at any point of time t for call admission control. AST of node n specifies the strength of radio connection of its neighbors with respect to n at a particular direction for call admission control. Angle - SINR Table for node n time t is shown below (Table I) where. we assume that nodes I, j and k are the neighbors of n. [9,10,11,12]

TABLE I.        ANGLE – SINR TABLE (AST) FOR NODE n

| Azimuth Angle (degree) | SINR value as perceived by neighbors of rooters n at different angle w.r.t rooters n | | |
|---|---|---|---|
| | i | j | K |
| 0 | SINR$^0$n,i$^{(t)}$ | SINR$^0$n,j$^{(t)}$ | SINR$^0$n,j$^{(t)}$ |
| 30 | SINR$^{30}$n,i$^{(t)}$ | SINR$^{30}$n,j$^{(t)}$ | SINR$^{30}$n,j$^{(t)}$ |
| 60 | SINR$^{60}$n,i$^{(t)}$ | SINR$^{60}$n,j$^{(t)}$ | SINR$^{60}$n,j$^{(t)}$ |
| … | ….. | ….. | …. |
| 330 | SINR$^{330}$n,i$^{(t)}$ | SINR$^{330}$n,j$^{(t)}$ | SINR$^{330}$n,j$^{(t)}$ |
| 360 | SINR$^{360}$n,i$^{(t)}$ | SINR$^{360}$n,j$^{(t)}$ | SINR$^{360}$n,j$^{(t)}$ |

In order to form ANGLE – SINR TABLE (AST), each node periodically sends a directional request in the form of a directional broadcast for the call admission control, sequentially in all direction. In this work, it has been done at 30 degree interval, covering the entire 360 degree space sequentially. A node is i in the neighborhood of n will wait until it receives all the request packets generated by n in all direction in that occasion. In other words, node I accumulates the entire column of the AST of n for node I, I accumulates the entire column of the AST of n for rooters i. Here, rooters i, after receiving the first request from n, has to wait a pre-specified amount of time to make sure that the directional broadcasts by n in all direction are over. Rooters I sends this information from all the neighbors of n, the Angle-SINR Table of n would be complete.[13]

## CONCLUSION

In this paper, we consider Call Admission Control for Traffic Management (CACTM) in Mobile Networks. Through the mathematical analysis and also present outage probability and a system model's for CAC. We also present an example for Call Admission Control for Traffic Management (CACTM) and present a frame work for the set up-the call admission control.

## ACKNOWLEDGMENT

The author would like to thank Dr. H.N. Dutta, Director, REMTech for his moral support in carrying out the work.

## AUTHORS PROFILE

Authors Profile .. K K Gautam is the Dean in the Roorkee Engineering & Management Technology Institute, Shamli-247 774, India.

Anurag Rai is the Head of the Information Technology Department in College of Engineering Roorkee,Roorkee-247 667, India.